\documentclass[onecolumn,hideversion]{cdcarticle}

\usepackage{amsfonts}
\usepackage{amssymb}
\usepackage{amsmath}
\usepackage{amsthm}
\usepackage{stdmath}
\usepackage{graphicx}

\graphicspath{{graphics_included/}}

\def\bk{\hspace{-0.75mm}}

\def\X{\mathcal{X}}

\begin{document}

\title{A Delay Analysis of Maximal Matching Switching with Speedup}

\author{%
  Randy Cogill\footnotesymbol{1}
  \and
  Sanjay Lall\footnotesymbol{2}
  }

\note{}

\maketitle

%%%%%%%%%%%%%%%%%%%%%%%%%%%%%%%%%%%%%%%%%%%%%%%%%%%%%%%%%%%%%%%%%%%%%%%%%%%%%%
% the footnotes

\makefootnote{1}{Department of Electrical Engineering,
  Stanford University, Stanford, CA 94305, U.S.A.\\
  Email: rcogill@stanford.edu}

\makefootnote{2}{%
  Department of Aeronautics and Astronautics,
  Stanford University, Stanford CA 94305-4035, U.S.A.\\
  Email lall@stanford.edu}

\makefootnote{1}{The first author was partially supported
  by a Stanford Graduate Fellowship.}

\makefootnote{1,2}{Partially supported by the Stanford URI
  \emph{Architectures for Secure and Robust Distributed
    Infrastructures}, AFOSR DoD award number 49620-01-1-0365.  }

%%%%%%%%%%%%%%%%%%%%%%%%%%%%%%%%%%%%%%%%%%%%%%%%%%%%%%%%%%%%%%%%%%%%%%%%%%%%%%

\begin{abstract}

In this paper we analyze the average queue lengths in a combined input-output queued switch using a maximal size matching scheduling algorithm. We compare these average queue lengths to the average queue lengths achieved by an optimal switch. We model the cell arrival process as independent and identically distributed between time slots and uniformly distributed among input and output ports. For switches with many input and output ports, the backlog associated with maximal size matching with speedup $3$ is no more than $3\frac{1}{3}$ times the backlog associated with an optimal switch. Moreover, this performance ratio rapidly approaches $2$ as speedup increases. 

\end{abstract}

%%%%%%%%%%%%%%%%%%%%%%%%%%%%%%%%%%%%%%%%%%%%%%%%%%%%%%%%%%%%%%%%%%%%%%%%%%%%%%

\section{Introduction}

Although packet switches vary in their internal construction, the most common architecture for high performance switches is the \emph{crossbar} switch. A crossbar switch contains $N$ input lines and $N$ output lines, where each input line meets each output line at a \emph{crosspoint}. This is depicted in Figure 1. When a crosspoint connecting and input line and an output line is closed, cells may be transferred between this input and output. Crossbar switches operate with the constraint that, when routing cells from inputs to outputs, each input may only be connected to a single output, and each output may only be connected to a single input. 

\vspace{2mm}

\begin{figure}[ht!]
\label{fig:crossbar}
\begin{center}
\scalebox{0.4}{\includegraphics{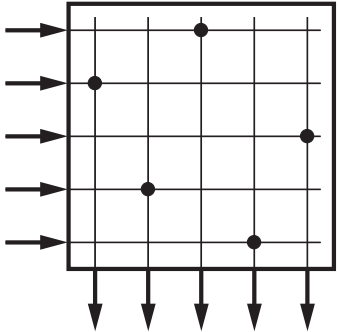}}
\caption{A depiction of a crossbar switch, where input lines are connected to output lines by closing crosspoints.}
\end{center}
\end{figure}

Switches are generally analyzed under a model where time is slotted, and only one cell may arrive at each input and depart from each output per time slot. Each arriving cell has a destination output port to which it must eventually be sent. Since multiple cells with the same output destination may arrive simultaneously at the input ports, switches require some form of buffering to store the cells which can not be immediately output. Buffered crossbar switches vary in their architecture, with the simplest being the output queued switch. In an output queued switch, buffers are placed at each output. All arriving cells are placed in their respective output queues in each time slot, and the output queues are served on a first come-first served (FCFS) basis. The average queue backlogs achieved by output queueing are minimum among all buffered crossbar switches. However, for a switch with $N$ inputs and $N$ outputs, output queueing requires that the as many as $N$ rounds of scheduling must be performed by the switch in each time slot (consider the case when $N$ cells destined for a single output arrive simultaneously). It is this requirement that makes output queueing infeasible for switches with many input and output ports.

Another alternative is the input queued switch, where cells are placed in buffers at the input ports in which they arrive. Delay performance of an input queued switch is heavily dependent on the service discipline used to serve the queues. It was shown in \cite{Anderson93} that if input queues are served FCFS, the switch can only achieve $58\%$ throughput. That is, suppose cells destined for output $j$ arrive at input $i$ at an average rate of $\lambda/N$ per time slot for all $i,j$. Then average backlogs are bounded for all $\lambda<1$ under output queueing, but average backlogs are only bounded for all $\lambda<0.58$ when input queues are served FCFS. However, if we use a service discipline which schedules cells in each input queue based on their destinations, it is possible to achieve $100\%$ throughput with input queueing. In particular, it was shown in \cite{McKeown96} that $100\%$ throughput is achieved by using a service discipline based on constructing maximum weight matchings (MWM) between inputs and outputs in each round of scheduling. This has the advantage over output queueing that only one round of scheduling is required per time slot. However, the algorithms required for computing maximum weight matchings are computationally expensive to implement. Also, the only known bounds on the average backlog under MWM are $O(N^2)$ \cite{Leonardi03}, as opposed to output queueing which has an average backlog which increases as $O(N)$.

Combined input-output queued (CIOQ) switches are an alternative to purely input queued or purely output queued switches. Combined input-output queued switches place buffers at both the input ports and the output ports, and perform some moderate number $s\ll N$ rounds of scheduling per time slot. The number of rounds of scheduling $s$ is commonly referred to the \emph{speedup} of the switch. It was shown in \cite{Dai00} that $100\%$ throughput can be achieved by using speedup $s=2$ and a simple service discipline based on greedily constructing maximal size matchings between input ports and output ports in each round of scheduling. Unlike pure input queueing with MWM scheduling, maximal size matching schedules can be computed with low computational cost. Also, unlike pure output queueing, the speedup requirements do not increase with the size of the switch.

The purpose of this paper is to show that average backlog performance of a CIOQ switch using maximal matching scheduling with low speedup is comparable to that of an output queued switch. Several previous papers have addressed the problem of analyzing backlogs in combined input-output queued switches with speedup. In \cite{Chuang99}, it was shown that under any traffic, an output queued switch can be \emph{exactly} emulated by a CIOQ switch operation with speedup $2$. However, the queueing discipline used in each round of scheduling has quite high computational cost. In \cite{Leonardi03}, an upper bound on average backlog was proven for maximal matching scheduling with speedup 2 assuming IID Bernoulli traffic with uniform loading on input and output ports. Unlike the best known bound for MWM scheduling, the ratio between this upper bound and a lower bound on the backlog for an output queued switch is constant as $N$ increases. However, this ratio becomes arbitrarily large as the arrival rate $\lambda$ approaches $1$. The same problem was considered and another upper bound on backlog was computed in \cite{Shah03}. There it was shown that the average backlog associated with maximal matching with speedup 2 is no more than $5$ times the backlog associated with an output queued switch. In this paper we also consider switches under uniformly loaded IID traffic. We show that average backlog associated with maximal matching with speedup $s$ gets arbitrarily close to $2$ times the backlog associated with an output queued switch as $s$ increases. Specifically, for a switch with many input and output ports, we show that for for speedup $s=3$, the backlog associated with maximal matching with speedup $3$ is no more than $3\frac{1}{3}$ times the backlog associated with an output queued switch. This performance ratio rapidly approaches $2$ as $s$ increases.

%%%%%%%%%%%%%%%%%%%%%%%%%%%%%%%%%%%%%%%%%%%%%%%%%%%%%%%%%%%%%%%%%%%%%%%%%%%%%%

\section{Preliminaries}

\subsection{Maximal Size Matchings}

Performing a round of scheduling in a crossbar switch can be thought of as constructing a matching in a bipartite graph $G$. This is shown in Figure 2. The vertices in $G$ represent input and output ports, and there is an edge between vertices $i$ and $j$ if the queue at input port $i$ contains a cell to be sent to output port $j$. Scheduling corresponds to choosing a collection of edges in $G$. That is, edge $(i,j)$ is chosen if a cell is to be sent from input port $i$ to output port $j$. The connectivity constraint imposed by the crossbar requires that the scheduled transfers correspond to a matching in the graph. A matching is a subgraph of $G$ with the defining property that no two edges are incident on the same vertex.

\vspace{2mm}

\begin{figure}[ht!]
\label{fig:matching}
\begin{center}
\scalebox{0.6}{\includegraphics{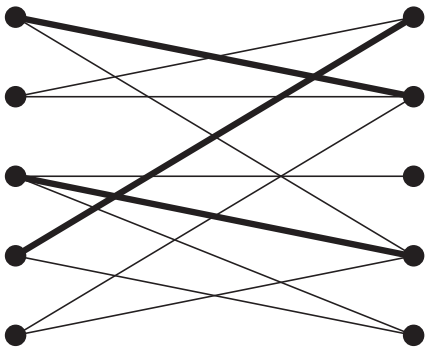}}
\caption{A bipartite graph, where the heavy lines show a maximal size matching.}
\end{center}
\end{figure}

Scheduling algorithms for input and combined input-output queued switches essentially amount to various criteria for selecting matchings. In this paper we consider \emph{maximal size matchings}. The main advantage to scheduling using maximal size matchings is that these matchings can be computed very efficiently using a simple greedy algorithm. A maximal size matching is a subgraph $H \subset G$ with the property that if we add any edge in $G-H$ to $H$, then $H$ is no longer a matching. The key property of maximal size matchings which is used in our later proofs is that if edge $(i,j)$ is in $G$, then there is an edge in $H$ incident to either vertex $i$ or vertex $j$.

\subsection{The Markov Chain Switch Model}

Here we assume a traffic model in which at most one cell may arrive at each input in a single time slot, and that cell arrivals at all time slots are independent and identically distributed. We let $A_{ij}(t) \in \{0,1\}$ be the random variable giving the number of cells arriving at input $i$ destined for output $j$ in time slot $t$. For simplicity, here we consider the case where arrivals are uniformly and independently distributed across inputs and outputs. This implies that the first and second moments of $A_{ij}(t)$ are
\begin{eqnarray*}
E[A_{ij}(t)] &=& \frac{\lambda}{N} \qquad \text{for all } i,j \\ 
E[A_{ij}(t)^2] &=& \frac{\lambda}{N} \qquad \text{for all } i,j \\
E[A_{kj}(t)A_{lj}(t)] &=& \frac{\lambda^2}{N^2} \quad \text{for all } j \text{ and all } k\ne l \\
E[A_{ik}(t)A_{il}(t)] &=& 0 \qquad \text{for all } i \text{ and all } k\ne l,
\end{eqnarray*}
where $0 \le \lambda < 1$ is a parameter describing the traffic intensity. Let $D_{ij}(t) \in \{0,\ldots,s\}$ denote the number of cells sent from input queue $i$ to output queue $j$ in time slot $t$, and let $E_j(t) \in \{0,1\}$ be the number of cells served from output queue $j$ in time slot $t$. Also, we let $X_{ij}(t)$ denote the number of cells in input queue $i$ destined for output $j$ in time slot $t$, and let $Y_j(t)$ denote the number of cells in output queue $j$ in time slot $t$. These random variables satisfy
\begin{eqnarray*}
X_{ij}(t+1) &=& X_{ij}(t) + A_{ij}(t) - D_{ij}(t) \\
Y_j(t+1) &=& Y_j(t) + \sum_{i=1}^ND_{ij}(t) - E_j(t).
\end{eqnarray*}
Throughout this paper, we will occasionally write these quantities in lowercase when simply referring to feasible values that they may take.

We will consider the problem of controlling the system to regulate the steady-state average per-period backlog in the input and output queues,
\[
\lim_{t\rightarrow\infty}\frac{1}{t+1}\sum_{k=0}^t\bk E\bk\left[\bk\sum_{i=1}^N\sum_{j=1}^NX_{ij}(k)\bk +\bk\bk \left. \sum_{j=1}^NY_j(k)\right|\bk X(0),Y(0)\bk\right].
\]
Under a maximal matching scheduling policy, $D(t)$ and $E(t)$ depend only on $X(t)$ and $Y(t)$. When this is the case, this system evolves as a Markov chain and we can use the following lemma to bound the average per-period backlog. This lemma is a special case of a more general result shown in \cite{Cogill06}. Results similar to the lemma below also appear, for example, in \cite{Meyn93}.

\begin{lem}
\label{lem:markov_bounds}
Consider a Markov chain $X$ with state space $\X$. For the cost function $r:\X\rightarrow\mathbb{R}$, let
\[
J(z) = \lim_{t\rightarrow\infty}\frac{1}{t+1}\sum_{k=0}^tE[r(X(k))|X(0)=z].
\] 
For any $h_U:\X\rightarrow\mathbb{R}$ such that $\displaystyle{\inf_{x\in\X}\{h_U(x)\} > -\infty}$,
\[
J(z) \le \sup_{x\in\X}\big\{r(x)\bk+\bk E[h_U(X(t\bk+\bk 1))|X(t)\bk=\bk x]-h_U(x)\big\}\]
for all $z\in \X$.
\end{lem}

\vspace{2mm}

\begin{proof} 
Let  $y = \inf_{x\in\X}\{h_U(x)\}$,
\[
\Delta_U(x) = E[h_U(X(t+1))|X(t)\bk=\bk x]-h_U(x),
\]
and
\[
\beta_U = \sup_{x\in\X}\big\{r(x)+\Delta(x)\big\}.
\]
For all $t\ge 0$,
\begin{eqnarray*}
\frac{1}{t+1}\sum_{k=0}^t E[r(X(k))|X(0)] &=& \frac{1}{t+1}\sum_{k=0}^tE[r(X(k)) + \Delta_U(X(k))|X(0)] \\
&& +\frac{h_U(X(0)) - y}{t+1} - \frac{E[h_U(X(t + 1))|X(0)] - y}{t+1} \\
&\le& \beta_U+\frac{h_U(X(0))-y}{t+1}.
\end{eqnarray*}
Clearly
\[
\lim_{t\rightarrow\infty}\frac{h_U(X(0))-y}{t+1}  = 0
\]
for all $z\in \X$, hence 
\[
\lim_{t\rightarrow\infty}\frac{1}{t+1}\sum_{k=0}^t E[r(X(k))|X(0)] \le \beta_U.
\]
\end{proof}

%%%%%%%%%%%%%%%%%%%%%%%%%%%%%%%%%%%%%%%%%%%%%%%%%%%%%%%%%%%%%%%%%%%%%%%%%%%%%%

\section{Main Result}

Our overall goal is to show that maximal matching scheduling with a speedup of $s$ keeps the average backlog are relatively close to the backlog achieved by an output queued switch. Specifically, we will: (i) compute an upper bound on the backlog associated with maximal matching with speedup $s$, (ii) compute the backlog associated with an output queued switch, and (iii) compute a bound on the ratio of these quantities.

Our first step will be to compute the upper bound on the average backlog. The following lemma will be used in the proof of the bound. Here we will define the quantities
\begin{eqnarray*}
\alpha_1(s) &=& \frac{1}{(s-1)^2-1}, \\
\alpha_2(s) &=& \frac{s-1}{(s-1)^2-1},
\end{eqnarray*}
and
\[
Q_{ij}(\lambda,d,e,s) = (1+\alpha_1(s)+\alpha_2(s))\lambda - \left(\alpha_1(s)\sum_{l=1}^Nd_{lj} + e_j + \alpha_2(s)\sum_{l=1}^Nd_{il}\right)
\]
which will be used throughout the rest of this paper.

\begin{lem}
\label{lem:stability}
For a CIOQ switch operating at speedup $s\ge 3$ using maximal matching scheduling, 
\[
Q_{ij}(\lambda,d,e,s)x_{ij} \le (\lambda-1)x_{ij}
\] 
for all $\lambda \le 1$ and all feasible values of $x_{ij}$, $d$, and $e$.
\end{lem}

\vspace{2mm}

\begin{proof}
When operating at speedup $s$, $s$ rounds of scheduling occur in each time slot. When using maximal matching scheduling, if there is a cell in input queue $i$ destined for output $j$ at the start of a round of scheduling, then either a cell is removed from input $i$ or a cell is sent to output $j$ in that round.
Also, if a cell is sent to output queue $j$, then output queue $j$ is served at the end of the time slot.

It is clear that the lemma holds if  $x_{ij}=0$. To prove the lemma for $x_{ij}>0$, we will consider three cases:
\begin{enumerate}
\item When $\sum_{l=1}^Nd_{lj} = s$ and $\sum_{l=1}^Nd_{il} = 0$,
\begin{eqnarray*}
\alpha_1(s)\sum_{l=1}^Nd_{lj} + e_j + \alpha_2(s)\sum_{l=1}^Nd_{il} &=& \alpha_1(s)s + 1 \\
&=& \frac{s-1+1}{(s-1)^2-1}+1 \\
&=& 1+\alpha_1(s)+\alpha_2(s).
\end{eqnarray*}

\item When $\sum_{l=1}^Nd_{lj} = 0$ and $\sum_{l=1}^Nd_{il} = s$,
\begin{eqnarray*}
\alpha_1(s)\sum_{l=1}^Nd_{lj} + e_j + \alpha_2(s)\sum_{l=1}^Nd_{il} &\ge& \alpha_2(s)s \\
&=& \frac{(s-1)(s-1+1)-1+1}{(s-1)^2-1} \\
&=& 1+\alpha_1(s)+\alpha_2(s).
\end{eqnarray*}

\item When $\sum_{l=1}^Nd_{lj} > 0$ and $\sum_{l=1}^Nd_{il} > 0$, it is sufficient to consider the case where $\sum_{l=1}^Nd_{lj} = \sum_{l=1}^Nd_{il} = 1$ since $\alpha_1(s) \ge 0$ and $\alpha_2(s) \ge 0$. In this case,
\begin{eqnarray*}
\alpha_1(s)\sum_{l=1}^Nd_{lj} + e_j + \alpha_2(s)\sum_{l=1}^Nd_{il} = 1+\alpha_1(s)+\alpha_2(s).
\end{eqnarray*}

\end{enumerate}

\noindent Note that if $x_{ij} > 0$ and 
\[
\sum_{l=1}^Nx_{lj} + \sum_{l=1}^Nx_{il} - x_{ij} < s,
\]  
then the total number of cells either in input queue $i$ or destined for output queue $j$ is less than $s$. However, in this case at least one cell must be sent from input queue $i$ to output queue $j$, implying that  $\sum_{l=1}^Nd_{lj} > 0$ and $\sum_{l=1}^Nd_{il} > 0$.
\end{proof}

\vspace{2mm}

\noindent Now we are ready to prove the upper bound on the backlog associated with maximal matching scheduling. We will let $J_\text{MMs}$ denote the average per-period backlog associated with maximal matching scheduling with speedup $s$.

\begin{thm}
\label{thm:upper_bound}
A CIOQ switch operating with speedup $s$ using a maximal matching scheduling policy has average per-period backlog satisfying
\[
J_{MMs} \le \frac{\left(k_1(s)\left(1-\frac{1}{N}\right)\lambda^2 + k_2(s)\lambda- k_3(s)\lambda^2\right)N}{2(1-\lambda)},
\]
where
\begin{eqnarray*}
k_1(s) &=& 1+\alpha_1(s) \\
k_2(s) &=& 2+(\alpha_1(s)+\alpha_2(s))(s+1) \\
k_3(s) &=& 2+2\alpha_1(s)+2\alpha_2(s).
\end{eqnarray*}
\end{thm}

\vspace{2mm}

\begin{proof} 
We prove this bound using Lemma \ref{lem:markov_bounds} with 
\[
h_U(x,y) = h_1(x)+h_2(x)+h_3(x,y),
\] 
where
\[
h_1(x) \bk=\bk \frac{\alpha_1(s)}{2(1-\lambda)}\sum_{j=1}^N\left(\bk\bk\left(\sum_{i=1}^Nx_{ij}\right)^2 + (s-2\lambda)\sum_{i=1}^Nx_{ij}\bk\right)
\]
\[
h_2(x) \bk=\bk \frac{\alpha_2(s)}{2(1-\lambda)}\sum_{i=1}^N\left(\bk\bk\left(\sum_{j=1}^Nx_{ij}\right)^2 + (s-2\lambda)\sum_{j=1}^Nx_{ij}\bk\right)
\]
\[
h_3(x,y)\bk=\bk \frac{1}{2(1-\lambda)}\sum_{j=1}^N\left(\sum_{i=1}^Nx_{ij}\bk+\bk y_j\right)^2 + \frac{1-2\lambda}{2(1-\lambda)}\bk\sum_{j=1}^N\left(\sum_{i=1}^Nx_{ij}\bk+\bk y_j\right)
\]
Since $h_U$ is quadratic with positive second order coefficients, it is clear that 
\[
\inf_{x,y}\{h_U(x,y)\} > -\infty, 
\]
satisfying the required condition of Lemma \ref{lem:markov_bounds}. Let 
\[
\Delta_i(x,y,d) = E[h_i(X(t+1),Y(t+1)|x,y,d] - h_i(x,y)
\]
denote the expected drift in $h_i$ when in state $(x,y)$ and action $d$ is taken.
\begin{eqnarray*}
\Delta_1(x,y,d)\bk\bk &=&\bk\bk \frac{\alpha_1(s)}{1-\lambda}\sum_{j=1}^N\left(\bk E\bk\left[\sum_{l=1}^N(A_{lj}-d_{lj})\right]\right)\sum_{i=1}^Nx_{ij} \\ && + \frac{\alpha_1(s)}{2(1-\lambda)}\sum_{j=1}^NE\left[\left(\sum_{l=1}^N(A_{lj}-d_{lj})\right)^2\right] \\
&& + \frac{\alpha_1(s)(s-2\lambda)}{2(1-\lambda)}\sum_{j=1}^NE\bk\left[\sum_{l=1}^N(A_{lj}-d_{lj})\right] \\
&\le& \bk\bk\frac{\alpha_1(s)}{1-\lambda}\sum_{i=1}^N\sum_{j=1}^N\left(\lambda-\sum_{l=1}^Nd_{lj}\right)x_{ij} + \alpha_1(s)\frac{\left(\left(1-\frac{1}{N}\right)\bk\lambda^2\bk +\bk (s\bk +\bk 1)\lambda\bk -\bk 2\lambda^2\right)\bk N}{2(1-\lambda)},
\end{eqnarray*}
where we used the fact that
\begin{eqnarray*}
E\left[\left(\sum_{l=1}^NA_{lj}\right)^2\right] &=& \sum_{k=1}^N\sum_{l=1}^NE[A_{kj}A_{lj}] \\
&=& \left(1 - \frac{1}{N}\right)\lambda^2 + \lambda.
\end{eqnarray*}
Similarly,
\begin{eqnarray*}
\Delta_2(x,y,d)\bk\bk &=&\bk\bk \frac{\alpha_2(s)}{1-\lambda}\sum_{i=1}^N\left(\bk E\bk\left[\sum_{l=1}^N(A_{il}-d_{il})\right]\right)\sum_{j=1}^Nx_{ij} \\
&& + \frac{\alpha(2)}{2(1-\lambda)}\sum_{i=1}^NE\left[\left(\sum_{l=1}^N(A_{il}-d_{il})\right)^2\right] \\
&& + \frac{\alpha_2(s)(s-2\lambda)}{2(1-\lambda)}\sum_{i=1}^NE\bk\left[\sum_{l=1}^N(A_{il}-d_{il})\right] \\
&\le& \bk\bk\frac{\alpha_2(s)}{2(1-\lambda)}\sum_{i=1}^N\sum_{j=1}^N\left(\lambda-\sum_{l=1}^Nd_{il}\right)x_{ij} + \alpha_2(s)\frac{\left((s+1)\lambda - 2\lambda^2\right)N}{2(1-\lambda)},
\end{eqnarray*}
where we used the fact that 
\begin{eqnarray*}
E\left[\left(\sum_{l=1}^NA_{il}\right)^2\right] &=& \sum_{k=1}^N\sum_{l=1}^NE[A_{ik}A_{il}] \\
&=& \lambda.
\end{eqnarray*}
Also,
\begin{eqnarray*}
\Delta_3(x,y)\bk\bk &=&\bk\bk \frac{1}{1-\lambda}\sum_{j=1}^N\bk\left(\bk E\bk\left[\sum_{l=1}^N\bk A_{lj}\bk -\bk e_j\bk\right]\bk \right)\bk\left(\bk\sum_{i=1}^N\bk x_{ij}\bk +\bk y_j\bk\right) \\
&& + \frac{1}{2(1-\lambda)}\sum_{j=1}^NE\left[\left(\sum_{l=1}^NA_{lj}-e_j\right)^2\right] \\
&& + \frac{1-2\lambda}{2(1-\lambda)}\sum_{j=1}^NE\left[\sum_{l=1}^NA_{lj}-e_j\right] \\
&=& \frac{1}{1-\lambda}\sum_{j=1}^N(\lambda-e_j)\left(\sum_{i=1}^Nx_{ij}+y_j\right) + \frac{\left(\left(1-\frac{1}{N}\right)\lambda^2 + 2\lambda- 2\lambda^2\right)N}{2(1-\lambda)}.
\end{eqnarray*}
Therefore,
\begin{eqnarray*}
r(x,y,d) + \sum_{i=1}^3\Delta_i(x,y,d) &\le& \sum_{i=1}^N\sum_{j=1}^N \left(1+\frac{Q_{ij}(\lambda,d,e,s)}{1-\lambda}\right)x_{ij} \\
&& + \sum_{j=1}^N\left(1+\frac{\lambda-e_j}{1-\lambda}\right)y_j \\
&& + \frac{\left(k_1(s)\left(1-\frac{1}{N}\right)\lambda^2 + k_2(s)\lambda- k_3(s)\lambda^2\right)N}{2(1-\lambda)}.
\end{eqnarray*}
From Lemma~\ref{lem:stability}, we have 
\[
\sum_{i=1}^N\sum_{j=1}^N \left(1+\frac{Q_{ij}(\lambda,d,e,s)}{1-\lambda}\right)x_{ij} \le 0
\]
for all values of $x$. Also, since $e_j=1$ if $y_j > 0$,
\[
\sum_{j=1}^N\left(1+\frac{\lambda-e_j}{1-\lambda}\right)y_j = 0
\]
for all values of $y$. Therefore,
\begin{eqnarray*}
J_{MMs}\bk\bk &=&\bk\bk \sup_{x,y}\{r(x,y) + \Delta(x,y,d)\} \\
&\le& \bk\bk \frac{\left(k_1(s)\left(1-\frac{1}{N}\right)\lambda^2 + k_2(s)\lambda- k_3(s)\lambda^2\right)N}{2(1-\lambda)}.
\end{eqnarray*}
\end{proof}

\vspace{2mm}

\noindent The previous theorem established an upper bound on $J_\text{MMs}$, the average per-period backlog under maximal matching with some fixed speedup $s$. We would now like to determine the expected per-period backlog associated with an output queued switch, which we will denote by $J_\text{OQ}$. This is a standard result, but is presented here to keep our treatment self-contained.

\begin{lem}
\label{thm:lower_bound}
An output queued switch has
\[
J_\text{OQ} = \frac{\left(\left(1-\frac{1}{N}\right)\lambda^2 +2\lambda - 2\lambda^2\right)N}{2(1-\lambda)}.
\]
\end{lem}

\vspace{2mm}

\begin{proof} 
Output queue $j$ is a discrete-time queue with queue with arrival process $A_{1j}+\cdots+A_{Nj}$. By the Pollaczek-Khintchine formula, (see, for example, \cite{Gross85}) the average steady-state per-period backlog of output queue $j$ is
\[
\frac{E\left[\left(\sum_{i=1}^NA_{ij}(t)\right)^2\right]+\lambda-2\lambda^2}{2(1-\lambda)}.
\]
Using the fact that
\begin{eqnarray*}
E\left[\left(\sum_{l=1}^NA_{lj}\right)^2\right] &=& \sum_{k=1}^N\sum_{l=1}^NE[A_{kj}A_{lj}] \\
&=& \left(1 - \frac{1}{N}\right)\lambda^2 + \lambda,
\end{eqnarray*}
we sum over all output queues to obtain
\[
J_\text{OQ} = \frac{\left(\left(1-\frac{1}{N}\right)\lambda^2 + 2\lambda - \lambda^2\right)N}{2(1-\lambda)}.
\]
\end{proof}

\vspace{2mm}

\noindent The upper bound and the result of the previous lemma are now used to determine a bound on the performance ratio between maximal matching and output queueing.

\begin{thm}
\label{thm:uniform}
The ratio of the average backlog under maximal matching scheduling with speedup $s \ge 3$ to the average backlog of an output queued switch satisfies
\begin{eqnarray}
\label{eqn:gap}
\frac{J_\text{MMs}}{J_\text{OQ}} \le \frac{N}{N-1}\left(\frac{2(s-1)^2+(s-1)}{(s-1)^2-1}\right).
\end{eqnarray}
\end{thm}

\vspace{2mm}

\begin{proof}
From Theorem \ref{thm:upper_bound} and Theorem \ref{thm:lower_bound} we have
\[
\frac{J_\text{MMs}}{J_\text{OQ}} \le \frac{k_1(s)\left(1-\frac{1}{N}\right)\lambda+k_2(s)-k_3(s)\lambda}{2-\left(1+\frac{1}{N}\right)\lambda}.
\]
By differentiating, it is straightforward to show that for $s \ge 3$ and $N \ge 2$, the previous expression is increasing in $\lambda$ for $0 \le \lambda \le 1$. Therefore,
\begin{eqnarray*}
\frac{J_\text{MMs}}{J_\text{OQ}} &\le& \frac{k_1(s)\left(1-\frac{1}{N}\right)+k_2(s)-k_3(s)}{1-\frac{1}{N}} \\
&=& \frac{1+\alpha_1(s)s+\alpha_2(s)(s-1)-(1+\alpha_1(s))\frac{1}{N}}{1-\frac{1}{N}} \\
&\le& \frac{N}{N-1}(1+\alpha_1(s)s+\alpha_2(s)(s-1)) \\
&=& \frac{N}{N-1}\left(\frac{2(s-1)^2+(s-1)}{(s-1)^2-1}\right)
\end{eqnarray*}
\end{proof}

\vspace{2mm}

\noindent For large $N$, the performance ratio approaches $2$ as $s$ increases. Table \ref{tab:bound_vals} shows the value of this ratio for several low values of speedup.

\begin{table}[h]
\label{tab:bound_vals}
\begin{center}
\begin{tabular}{|c||c|c|c|c|c|}
\hline $s$ & $3$ & $4$ & $5$ & $8$ & $15$ \\ \hline
$J_\text{MMs}/J_\text{OQ}$ & $3.36$ & $2.65$ & $2.42$ & $2.20$ & $2.10$ \\ \hline
\end{tabular}
\end{center}
\caption{Values of (\ref{eqn:gap}) at several low values of speedup for a $128\times 128$ switch.}
\end{table}

%%%%%%%%%%%%%%%%%%%%%%%%%%%%%%%%%%%%%%%%%%%%%%%%%%%%%%%%%%%%%%%%%%%%%%%%%%%%%%

\section{Conclusions}

In this paper we have analyzed the average backlogs in network switches using a maximal size matching scheduling policy with speedup. It is shown that switches using maximal matching with speedup achieve backlogs comparable to an optimal switch. For the sake of simplicity, we have focused on the case of IID arrivals with uniform loading on input and output ports. We believe that the performance bounds proven in this paper can be tightened when arrivals are time correlated, and this is a subject of future research.

%%%%%%%%%%%%%%%%%%%%%%%%%%%%%%%%%%%%%%%%%%%%%%%%%%%%%%%%%%%%%%%%%%%%%%%%%%%%%%

\bibliography{switching}

\begin{thebibliography}{1}

\bibitem{Anderson93}
T.~Anderson, S.~Owicki, J.~Saxe, and C.~Thacker.
\newblock High speed switch scheduling for local area networks.
\newblock {\em ACM Trans. Comp. Sys.}, 11(4):319--351, 1993.

\bibitem{Chuang99}
S.T. Chuang, A.~Goel, N.~McKeown, and B.~Prabhakar.
\newblock Matching output queueing with a combined input-output queued switch.
\newblock {\em IEEE INFOCOM 1999}, 3:1169--1178, 1999.

\bibitem{Cogill06}
R.~Cogill and S.~Lall.
\newblock Suboptimality bounds in stochastic control: A queueing example.
\newblock {\em To appear in the Proceedings of the 2006 American Control
  Conf.}, 2006.

\bibitem{Dai00}
J.~Dai and B.~Prabhakar.
\newblock The throughput of data switches with and without speedup.
\newblock {\em IEEE INFOCOM 2000}, 2:556--564, 2000.

\bibitem{Gross85}
D.~Gross and C.~Harris.
\newblock {\em Fundamentals of Queueing Theory}.
\newblock John Wiley, New York, 1985.

\bibitem{Leonardi03}
E.~Leonardi, M.~Mellia, F.~Neri, and A.~Marsan.
\newblock Bounds on delays and queue lengths in input-queued cell switches.
\newblock {\em Journal of the ACM}, 50(4):520--550, 2003.

\bibitem{McKeown96}
N.~McKeown, V.~Anantharan, and J.~Walrand.
\newblock Achieving $100\%$ throughput in an input-queued switch.
\newblock {\em IEEE INFOCOM 1996}, 1:296--302, 1996.

\bibitem{Meyn93}
S.~Meyn and R.~Tweedie.
\newblock {\em Markov Chains and Stochastic Stability}.
\newblock Springer-Verlag, 1993.

\bibitem{Shah03}
D.~Shah.
\newblock Maximal matching scheduling is good enough.
\newblock {\em IEEE Globecom}, 22(1):3009--3013, 2003.

\end{thebibliography}

%%%%%%%%%%%%%%%%%%%%%%%%%%%%%%%%%%%%%%%%%%%%%%%%%%%%%%%%%%%%%%%%%%%%%%%%%%%%%%

\end{document}